\documentclass[aps,prb,superscriptaddress,showpacs,twocolumn,a4paper]{revtex4-2}
\usepackage{amsmath,amsthm,amssymb,bm}
\usepackage{float}
\usepackage[pdftex]{graphicx}
\usepackage[utf8]{inputenc}
\usepackage{xcolor}
\usepackage{mathtools}
\usepackage{ulem}
\providecommand{\includegraphics}[2][width=\textwidth]{$#2$}
\definecolor{citecol}{rgb}{0.152, 0.574, 0.664}
\definecolor{urlcol}{rgb}{0.152, 0.574, 0.664}
\usepackage[colorlinks=true, urlcolor=urlcol,citecolor=citecol]{hyperref}
\newcommand{\ua}{\uparrow}
\newcommand{\da}{\downarrow}

\begin{document}

\title{Chiral bosonic quantum spin liquid in the integer-spin Heisenberg-Kitaev model}
\author{Arnaud Ralko}
\affiliation{Institut N\'eel, Université Grenoble Alpes and CNRS, Grenoble, France}
\author{Jaime Merino}
\affiliation{Departamento de F\'isica Te\'orica de la Materia Condensada, Condensed Matter Physics Center (IFIMAC) and Instituto Nicol\'as Cabrera, Universidad Aut\'onoma de Madrid, Madrid 28049, Spain}

\date{\today}

\begin{abstract}
Motivated by the possibility of finding a bosonic quantum spin liquid in the integer spin-$S$ Heisenberg-Kitaev model on the honeycomb lattice, 
we derive a Schwinger boson mean field theory involving both singlet and triplet pairing channels which includes hopping and pairing operators
on equal footing. The mixed construction introduced here is justified by the good comparison with exact diagonalization energies of the
 $S \leq 3/2$ Heisenberg-Kitaev model and the perfect match with the Luttinger-Tisza semiclassical energies obtained at large-$S$. We find various 
competing gapped quantum spin liquids close to the Kitaev point. A comparison of their spin excitation spectrum 
with the dynamical structure factor obtained from exact diagonalizations allows us to identify the physical
spin liquid {\it Ansatz} of the model. In particular, we identify a chiral quantum spin liquid state whose spin excitation spectrum 
follows closely the exact diagonalization data and survives up to large spin $S \lesssim 2$. We propose this state as a promising 
quantum spin liquid candidate for the integer spin-$S$ antiferromagnetic Kitaev model which may be realized in $S=1$ Kitaev materials
A$_3$Ni$_2$XO$_6$ and KNiAsO$_4$.
\end{abstract}
\maketitle

\section{Introduction}

Quantum spin liquids are magnetically disordered states even at $T=0$ which are 
highly entangled and often display topological order and fractional excitations\cite{savary_quantum_2017}. A 
quantum spin liquid (QSL) whose spin excitations are fractionalized into Majorana fermions arises as the exact ground state of the
$S=1/2$ Kitaev spin model. Remarkably, non-Abelian anyons predicted in the Kitaev model \cite{kitaev_anyons_2006} may 
be detected through the half-quantization of the thermal Hall conductivity in the 
Kitaev candidate material $\alpha$-RuCl$_3$.\cite{kasahara_majorana_2018,yokoi_half-integer_2021}
Concomitantly with such great progress achieved on the $S=1/2$ Kitaev materials there is growing interest on the higher spin 
$S$ Kitaev model which is not exactly solvable. While A$_3$Ni$_2$XO$_6$ (A=Li, Na and X=Bi, Sb), 
Na$_2$Ni$_2$TeO$_6$ \cite{stavropoulos_microscopic_2019}, and KNiAsO$_4$ \cite{taddei_zigzag_2023} have been 
proposed as candidate materials of the $S=1$ Heisenberg-Kitaev model, the $S=3/2$  Kitaev model may be 
realized in Cr-based ferromagnets such as CrI$_3$,CrGeTe$_3$\cite{xu_interplay_2018}  
monolayers and  CrSiTe$_3$.\cite{xu_possible_2020}

Apart from the exact solution of the $S=1/2$ Kitaev model, the ground state of the classical $S \rightarrow \infty$ model  
is also well known and consists of a large number of degenerate cartesian states\cite{baskaran_spin-_2008,rousochatzakis_quantum_2018}. An order by disorder mechanism 
selects states with the smallest self-avoiding walks followed by the spins resulting in a fully packed hexagonal dimer plaquette 
configurations such as the one depicted in Fig.~\ref{fig:bss}. For intermediate spin values, $1/2<S<\infty$, however, the Kitaev 
model is not exactly solvable even though plaquette $\mathbb{Z}_2$ gauge fluxes can be 
constructed in terms of Majorana fermion operators commuting with the hamiltonian, as for $S=1/2$. However, the
general half-odd-integer Kitaev model with $S>1/2$ cannot be expressed in a simple way in terms of the Majorana 
operators \cite{baskaran_spin-_2008} as for $S=1/2$. In the integer spin case bosonic-type of excitations emerge instead 
in the system. Hence, it remains an open issue which is the nature of the ground state and the elementary excitations of the 
Kitaev model for arbitrary spin $S$.

Recent numerical works suggest that while the half-odd-integer Kitaev model hosts a QSL with Majorana 
fermion excitations similar to the $S=1/2$ case, the integer model is most probably gapped hosting excitations of 
the bosonic type\cite{lee_tensor_2020,khait_characterizing_2021,dong_spin-1_2020,zhu_magnetic_2020}. This is consistent with the conjecture based on 
the generalized parton construction \cite{ma_Z_2023} extended to $S >1/2$ 
which states that the ground state of the half-odd-integer $S$ model always consists of a deconfined $\mathbb{Z}_2$ gauge 
field coupled to giant Majorana fermions which leads to a gapless QSL with two giant Majorana cones.  
In contrast, the integer $S$ model is conjectured to host a $\mathbb{Z}_2$ gapped spin liquid with bosonic gauge charges.

In the present work we analyze the Heisenberg-Kitaev model with an arbitrary $S$ using 
a Schwinger boson approach. Our aim is to understand the physics of the integer $S$ Kitaev model in which bosonic 
excitations have been proposed to be the elementary spin excitations of the system as discussed above
justifying the use of this approach. Our Schwinger boson mean field theory (SBMFT) uses a mixed singlet/triplet representation 
based on all possible hopping and pairing operators constructed in these channels\cite{kos_quantum_2017,kargarian_unusual_2012,schneider_projective_2022}
rather than the more standard SU(2) approaches \cite{auerbach_interacting_1994,mezio_test_2011,mezio_low_2012,flint_symplectic_2009}. The reliability 
of the mean-field {\it Ans\"atze} used is established by comparing the SBMFT in the large-$S$ limit with the exact ground state of the 
classical model recovering the well known N\'eel, zig-zag, ferromagnetic, stripy magnetic orders  
for non-zero Heisenberg exchange, away from the pure Kitaev model. Remarkably, the SBMFT predicts 
different competing QSL solutions of the Kitaev model. A comparison of the 
SBMFT spin excitation spectra with exact diagonalization data allows us to single out the most likely 
QSL among all the {\it Ans\"atze} relevant to the integer-$S$ Kitaev model. An intricate 
chiral QSL is found to be the most robust solution surviving 
up to about  $S = 2$  hosting an excitation spectra in good agreement with exact diagonalization
data. We speculate that such mean-field state will be favored by quantum fluctuations beyond
the SBMFT approach becoming the true ground state of the $S=1$ Kitaev model. The possible existence 
of the $\mathbb{Z}_2$ gapped chiral QSL predicted here can be probed through 
experiments sensitive to time-reversal symmetry breaking on $S=1$ AF Kitaev materials such as
A$_3$Ni$_2$XO$_6$ and KNiAsO$_4$.

The rest of the paper is organized as follows. In Sec. \ref{sec:model} we introduce the Heisenberg-Kitaev model  and the SBMFT approach
used to analyze it. In Sec. \ref{sec:valid} we validate the SBMFT decoupling used here by comparison to exact diagonalizations (ED) calculations and semiclassical approaches. 
In Sec. \ref{sec:qsl} we discuss the QSL solutions found within SBMFT and their physical properties such as their spin excitation 
spectra are analyzed in Sec. \ref{sec:phys}. Finally, we end our work in Sec. \ref{sec:concl} with some conclusions discussing the implication of our results to experiments
on $S=1$ Kitaev materials.

\section{Model and methods}
\label{sec:model}
\begin{figure}[ht!] \centering
  \includegraphics[width=0.45\textwidth]{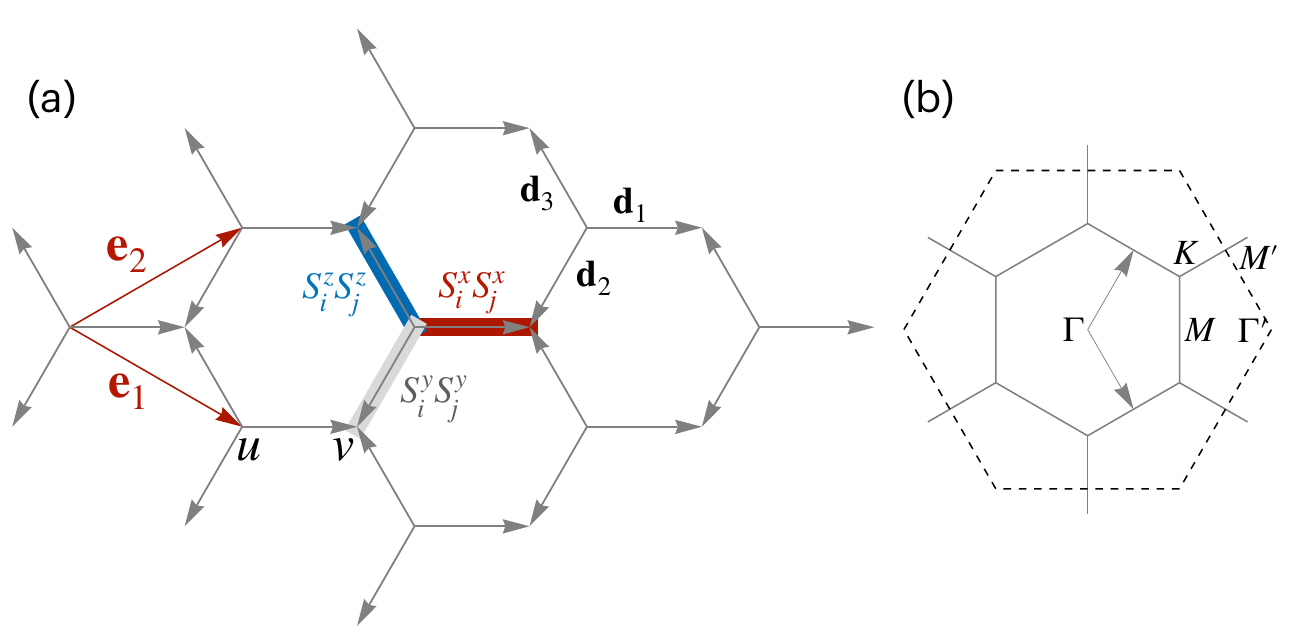}
  \caption{(a) The honeycomb lattice and its translation vectors $\vec{e}_1$ and $\vec{e}_2$, the two sub-lattices $u$ and $v$, the three bond dependent Kitaev couplings and the three nearest neighbor vectors $\{ \vec{d}_i \}$.   (b) The corresponding first (continuous lines) and second (dashed lines)  Brillouin zones, and the high symmetry points. For the dynamical structure factors, we consider the path $K\Gamma M \Gamma' M' K M$ to emphasize possible symmetry breakings.
  }
  \label{fig:lattice}
\end{figure}
{
The bipartite honeycomb lattice is defined by Bravais vectors  
\begin{eqnarray}
	{\bf e}_1 = \frac{a}{2}(\sqrt{3},-1)^T ,~~~{\bf e}_2 = \frac{a}{2}(\sqrt{3}, 1)^T,
\end{eqnarray}
with $a$ the lattice parameter set to 1 in the rest of the paper. As depicted in Fig.\ref{fig:lattice}, any sub-lattice $u$ has  3 nearest neighbors $v$ in directions:
\begin{eqnarray}
	{\bf d}_1 &=&  ({\bf e}_1+{\bf e}_2)/3, \nonumber \\
	{\bf d}_2 &=&  ({\bf e}_1-2{\bf e}_2)/3, \\
	{\bf d}_3 &=& ({\bf e}_2-2{\bf e}_1)/3. \nonumber 
\end{eqnarray}
}

The Heisenberg-Kitaev model on this lattice, with quantum spin ${\bf S}_i$ at site $i$, is given by the hamiltonian: 
\begin{eqnarray}
\label{ham}
{\cal H}_\text{HK} &=& \sum_{\langle i,j \rangle_\gamma} \hat{\bf S}_i   { J}^\gamma   \hat{\bf S}_j,
\end{eqnarray}
where $\gamma = 1,2,3$ is either the bond direction and the spin components $x, y, z$. In our definition, ${ J}^\gamma$ are the corresponding coupling diagonal matrices given by $\text{diag}( {J}^\gamma) = J_H(1,1,1)+J_K(\delta_{\gamma,1},\delta_{\gamma,2},\delta_{\gamma,3})$, where $\delta_{i,j}$ is the usual Kronecker delta, $J_H$ the isotropic Heisenberg interaction and $J_K$ the bond direction dependent Kitaev interaction which explicitly breaks the rotation symmetry of the lattice, as shown in Fig.~\ref{fig:lattice}(a).
In this work, we set 
\begin{eqnarray*}
	J_H = \cos \theta,~~~~ J_K= 2 \sin \theta,
\end{eqnarray*} 
and focus on  $\theta \in [ 0 , \pi ]$ for reasons that are explained hereafter. Note that thanks to the Klein duality \cite{chaloupka_zigzag_2013}, it is always possible to access corresponding ferromagnetic regimes.

Among the possible methods to study strongly correlated states of matter with possible disordered ground states such as QSLs, we 
consider two complementary quantum approaches, exact diagonalisation (ED) for systems with spin magnitudes,  $S \leq {3 \over 2} $, and a parton 
mean field theory. The latter has proven to be very versatile and successful. When partons are Schwinger bosons, it is possible to 
derive a mean field theory for pure Heisenberg models, which treats spin disordered states and magnetically ordered 
phases on equal footing. A key feature of such SBMFT is the representation in terms of bond operators  
that nicely connects to the expected Sp(N) symmetry group in the large-$N$ limit\cite{read_large-_1991,sachdev_kagome-_1992}. This feature 
will be used here to select the most relevant SBMFT by comparing its large-$S$ energies with the ones 
obtained through the Luttinger-Tisza approach (LTA) \cite{luttinger_note_1951,litvin_luttinger-tisza_1974} and ED at lower $S$.

We now want to recall the main lines of the SBMFT, more details can be found in [\onlinecite{merino_role_2018,lugan_topological_2019,lugan_schwinger_2022,halimeh_spin_2016,kos_quantum_2017,schneider_projective_2022,kargarian_unusual_2012}] and references therein. 
In the Schwinger boson construction, the spin operator at, say, site $i$ is written in terms of bosons as: 
\begin{eqnarray}
\hat{\bf{S}}_i &=& \frac{1}{2} {\hat b}_{i}^\dagger {\boldsymbol \sigma} {\hat b}_{i}
\end{eqnarray}
where ${\hat b}_i^\dagger =  ({\hat b}_{i,\ua}^\dagger,{\hat b}_{i,\da}^\dagger)$ are bosonic creation operators and ${\boldsymbol \sigma} = \sigma^1 {\bf u}_x + \sigma^2 {\bf u}_y + \sigma^3 {\bf u}_z$ are the Pauli matrices. Anticipating notations, we also define $\sigma^0$ as the $2 \times 2$ identity matrix and define $\alpha=0,1,2,3$ as the index of the components. For such a mapping to realize physically the spin $S$ algebra, it is necessary to satisfy the boson constraint 
\begin{eqnarray}
\hat{n}_i =  {\hat b}_{i}^\dagger  {\hat b}_{i} = 2 S = \kappa,
\end{eqnarray}
{\it i.e.}  all unphysical states not having exactly $\kappa$ bosons per site have to be projected out. If this constraint is strictly respected, the parton construction is exact. Unfortunately, it is very difficult to treat the constraint exactly in practice, and  it is then done only on average by introducing Lagrange multipliers:
\begin{eqnarray}
\sum_i \lambda_i \left(   \hat{n}_{i } - \kappa \right),
\end{eqnarray}
to the Hamiltonian ${\cal H}_{\text{HK}}$.  {A benefit is that $S$ can now be viewed as an external parameter controlling the strength of the quantum fluctuations}.

In Eq.~(\ref{ham}),  ${\cal H}_{\text{HK}}$ contains both the SU(2) symmetric Heisenberg and the non-SU(2) Kitaev terms which can be rewritten in terms of the hopping $\hat{h}^\alpha$ and pairing operators $\hat{p}^\alpha$ with $\alpha = 0$ belonging to the singlet  and $\alpha = 1,2,3$ triplet channels. Explicitly,  the two terms read: 
\begin{eqnarray}
\hat{h}_{ij}^\alpha = \frac{1}{2}  {\hat b}_{i}^\dagger  \sigma^{\alpha} {\hat b}_{j} ,~~~~\hat{p}_{ij}^\alpha = \frac{i}{2}  {\hat b}_{i} \left( \sigma^\alpha  \sigma^2 \right) {\hat b}_{j }.
\end{eqnarray}
 At this stage, we have 8 different operators at our disposal and many possible combinations of them can be used to describe exactly ${\cal H}_{\text{HK}}$. 
 Following and adapting the compact and useful notations of [\onlinecite{schneider_projective_2022}], it is  possible to obtain general identities of the form
\begin{eqnarray}
\label{identities}
\hat {\bf S}_i  { J}^\gamma  \hat{\bf S}_j &=& : {\hat{\bf h}}_{ij}^\dagger H_{ij}^\gamma {\hat{\bf h}}_{ij}: +  {\hat{\bf p}}_{ij}^\dagger P_{ij}^\gamma {\hat{\bf p}}_{ij} + K_{ij},
\end{eqnarray}
where $:{ \cal \hat{O}}:$ is the normal order of the operator ${\cal \hat{O}}$ and 
where we have introduced the vectors of hopping  ${\hat{\bf h}}_{ij} = (\hat{h}_{ij}^0,\hat{h}_{ij}^1,\hat{h}_{ij}^2,\hat{h}_{ij}^3)^T$ and pairing  ${\hat{\bf p}}_{ij} = (\hat{p}_{ij}^0,\hat{p}_{ij}^1,\hat{p}_{ij}^2,\hat{p}_{ij}^3)^T$ operators, and their corresponding bond dependent $4 \times 4$ coupling  matrices $H_{ij}^\gamma$, $P_{ij}^\gamma$, functions of $J_H$ and $J_K$. The last term $K_{ij}$ is a constant to be determined in function of the chosen mean field theory.
 
 The Schwinger boson hamiltonian for our present Heisenberg-Kitaev model can be generically expressed as: 
\begin{eqnarray}
\label{sbmftham}
{\cal H}_{\text{HK}} &=&  \sum_{\langle i,j \rangle_\gamma} : {\hat{\bf h}}_{ij}^\dagger H_{ij}^\gamma {\hat{\bf h}}_{ij}: +  {\hat{\bf p}}_{ij}^\dagger P_{ij}^\gamma {\hat{\bf p}}_{ij} + K_{ij} \nonumber \\
&+& \sum_i \lambda_i ({\hat n}_i - \kappa).
\end{eqnarray}
 As said above, several  choices of $H_{ij}^\gamma$ and $P_{ij}^\gamma$ are possible, yielding different mean field theories that can drastically change the results. For example, in the case of $H_{ij}^\gamma = 0$, only pairing terms are present, but this usually leads to a bad ground state energy and/or dynamical structure factors\cite{mezio_test_2011,flint_symplectic_2009}. On the other hand, when  $P_{ij}^\gamma = 0$, only magnetic orders are accessible by the theory and  possible QSL states may be overlooked.

A reasonable strategy would be to preserve the SU(2) symmetry at $J_K=0$. At this Heisenberg point, ${\cal H}_{\text{HK}}$ is only written with ${\hat h}_0$ and ${\hat p}_0$, and properly recovers the Sp($\mathcal{N}$) symmetry group in the large-$\mathcal{N}$ expansion limit\cite{read_large-_1991,sachdev_kagome-_1992}. 

A natural choice is to take $\text{diag}(H_{ij}^\gamma )= J_H(1,0,0,0) + J_K(1,0,0,0)$, $\text{diag}(P_{ij}^\gamma) = - J_H(1,0,0,0) - J_K (0,\delta_{\gamma,1}, \delta_{\gamma,2}, \delta_{\gamma,3})$ and $K_{ij} = 0$, corresponding to describe the non-SU(2) Kitaev bond by the help of triplet operators only, and the isotropic SU(2) Heisenberg one by the singlet operator. Similar choice has been successfully employed in other systems, {\it e.g.} for an exchange Hamiltonian arising in the strong interaction limit of the  layered heavy transition metal oxide compound (Li,Na)$_2$IrO$_3$\cite{kargarian_unusual_2012}, on a Heisenberg-Kitaev model on the triangular lattice\cite{kos_quantum_2017} or more recently for a XXZ model on the pyrochlore\cite{schneider_projective_2022} lattice. {For the present model, similar choice of SBMFT does not recover the expected large-$S$ classical limit and in addition provides a false energy profile as a function of $\theta$. Moreover,} the isotropic term and the three Kitaev bonds are treated independently {by defining 4 operators} such that $\hat{\bf S}_i \hat{\bf S}_j = \sum_{\gamma} \hat{S}_i^\gamma \hat{S}_j^\gamma$ is not necessarily {satisfied.\cite{kos_quantum_2017}}
 
In this paper, we follow another direction by noticing that the isotropic term can be written as a sum of singlet and triplet projectors $\hat{\bf S}_i \hat{\bf S}_j = -\frac{3}{4} | s_{ij} \rangle \langle s_{ij} | + \frac{1}{4} \left[ | t^1_{ij} \rangle \langle t^1_{ij} |+ | t^2_{ij} \rangle \langle t^2_{ij} | + | t^3_{ij} \rangle \langle t^3_{ij} | \right]$ from one side, and that ideally a SBMFT should recover the classical solutions  when $S \to \infty$ from the other side (see the next section for a comparison with the LT method\cite{luttinger_note_1951,litvin_luttinger-tisza_1974}). We then define the following operator matrices 
\begin{eqnarray}
	 P^\gamma_{ij}  = -J_H  m_0- J_K m_\gamma,
\end{eqnarray}
$H_{ij}^\gamma = - P_{ij}^\gamma$ and $K_{ij} = 0$ and where we have introduced  $\text{diag}(m_0) = ( 1,-\frac{1}{3},-\frac{1}{3},-\frac{1}{3} )$, and $m_\gamma$ with $\gamma =1,2,3$ as $\text{diag}(m_1) = (+\frac{1}{3},+\frac{1}{3},-\frac{1}{3},-\frac{1}{3})$, $\text{diag}(m_2) = (+\frac{1}{3},-\frac{1}{3},+\frac{1}{3},+\frac{1}{3})$ and $\text{diag}(m_3)= (+\frac{1}{3},-\frac{1}{3},-\frac{1}{3},+\frac{1}{3})$, verifying the relation $m_0 = m_1 + m_2 + m_3$. It is now clear that in terms of operators, we  have the identities 
\begin{eqnarray}
	\hat{\bf S}_i \hat{\bf S}_j = :\hat{\bf h}_{ij}^\dagger m_0 \hat{\bf h}_{ij}: -  \hat{\bf p}_{ij}^\dagger m_0 \hat{\bf p}_{ij},\nonumber \\
	\hat{S}_i^\gamma \hat{S}_j^\gamma = :\hat{\bf h}_{ij}^\dagger m_\gamma \hat{\bf h}_{ij}: -  \hat{\bf p}_{ij}^\dagger m_\gamma \hat{\bf p}_{ij},
\end{eqnarray}
satisfying $\hat{\bf S}_i \hat{\bf S}_j = \sum_{\gamma} \hat{S}_i^\gamma \hat{S}_j^\gamma$ by construction. 
To study the HKM,  we perform the following mean-field decoupling
\begin{eqnarray*}
: {\hat{\bf h}}_{ij}^\dagger H_{ij}^\gamma {\hat{\bf h}}_{ij}: &\to&  {\hat{\bf h}}_{ij}^\dagger H_{ij}^\gamma {{\bf h}}_{ij} +  {{\bf h}}_{ij}^* H_{ij}^\gamma {\hat{\bf h}}_{ij} -  {{\bf h}}_{ij}^* H_{ij}^\gamma {{\bf h}}_{ij}, \nonumber \\ 
  {\hat{\bf p}}_{ij}^\dagger P_{ij}^\gamma {\hat{\bf p}}_{ij}  &\to& {\hat{\bf p}}_{ij}^\dagger P_{ij}^\gamma {{\bf p}}_{ij}  + {{\bf p}}_{ij}^* P_{ij}^\gamma {\hat{\bf p}}_{ij} - {{\bf p}}_{ij}^* P_{ij}^\gamma {{\bf p}}_{ij},
\end{eqnarray*}
where we have defined the expectation vectors  ${{\bf p}}_{ij} = \langle \phi_0 | {\hat{\bf p}}_{ij}   | \phi_0 \rangle $ and ${{\bf h}}_{ij} = \langle \phi_0 | {\hat{\bf h}}_{ij}   | \phi_0 \rangle $ calculated in the vacuum bosonic ground state $| \phi_0 \rangle$ for each oriented pair of interacting spins $(i \to j)$, that we inject in Eq.~\ref{sbmftham} to get the SBMFT ${\cal H}_\text{HK}^\text{MF}$. 
Although quite standard, we reproduce the main steps to solving such mean-field hamiltonian for completeness. 
In order to be compatible with all known magnetic orders encountered in the HKM\cite{chaloupka_zigzag_2013,janssen_phase_2022,consoli_heisenberg-kitaev_2020,dong_spin-1_2020,lee_tensor_2020}, we consider a large magnetic unit cell of $n_u$ sites containing $12 \times n_u $ complex mean-field parameters (8 per bond in the unit-cell). We have tried unit cells up to $24$ sites and found that all {\it Ans\"atze}  of this work were {described} with at most  {unit-cell containing 8 sites. It is thus enough to only consider  $n_u = 8$  in our SBMFT calculations.} ${\cal H}_\text{HK}^\text{MF}$ is self-consistently solved, starting from random mean-field parameters $\{ {{\bf p}}_{ij} , {{\bf h}}_{ij} \}$ and adjusting the set of $n_u$ independent -- the unit cell is translationally invariant-- Lagrange multipliers $\{ \lambda_i \}$ to satisfy the boson constraint. We then obtain the ground state free of bosons $| \phi_0 \rangle$ by diagonalizing the $4 n_u \times 4 n_u$ $q$-dependent Hamiltonians written in the Fourier space on a Brillouin zone of linear size $l$ containing $l \times l$  momenta ($N = n_u \times l \times l$ sites). The diagonalization is performed using a Cholesky decomposition\cite{toth_linear_2015}. A new set of mean-field parameters is then computed in $| \phi_0 \rangle$, and the procedure is repeated until convergence of the mean-field variables to a desired tolerance (typically $10^{-11}$). Our procedure is then derivative-free, which is easier to work with complex parameters. Importantly, we stress that $| \phi_0 \rangle$ is the boson vacuum at $T=0$ with a gap scaling like $\simeq 1/l$ for an ordered state\cite{lugan_schwinger_2022}. This means that the condensation only appears in the thermodynamic limit, unless the state in purely classical, {\it e.g.} the ferromagnetic state.

\section{Validating the SBMFT}
\label{sec:valid} 

\begin{figure}[ht!] \centering
  \includegraphics[clip,width=0.45\textwidth]{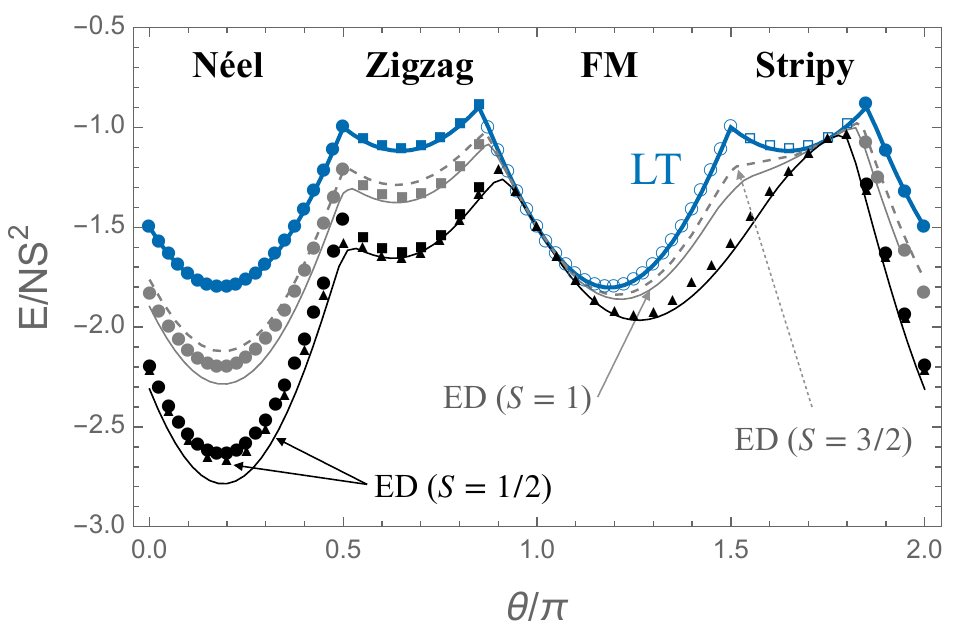}
  \caption{Ground state energies per site of the Heisenberg-Kitaev model  in function of $\theta$, obtained by ED on a 12 site cluster for $S=1/2$ (black line), $S=1$ (gray line) and $S=3/2$ (dashed gray line), and
  by  the Luttinger-Tisza method at the classical limit (blue line). Colored circles/squares are the corresponding SBMFT {\it Ans\"atze} containing only pairing/hopping operators. Empty symbols correspond to mean-field solutions reachable only by adding Bose condensates in the theory (beyond the scope of this work), such as for the non-fluctuating ferromagnetic state. {We show the ED energies on a 24 site cluster to illustrate  the smallness of the finite size effects in comparison with the SBMFT.} 
  }
  \label{fig:energy}
\end{figure}

We have calculated the SBMFT dependence of the ground state energy with the parameter $\theta \in [0,\pi]$ by solving self-consistently ${\cal H}_{\text{HK}}^{\text{MF}}$  for various $S$ which range from the strong quantum regime (small $S$) to the large-$S$ limit { recovering known results for  $S=1/2$\cite{chaloupka_zigzag_2013} and $S=1$.\cite{dong_spin-1_2020}} For the latter, the {\it Ans\"atze} can be obtained analytically allowing to fully span the parameter space $\theta \in [0,2\pi]$.
In Fig.~\ref{fig:energy}, we compare our results (symbols) with (i) the classical limit obtained by the Luttinger-Tisza method\cite{litvin_luttinger-tisza_1974} (blue line) and (ii) the exact diagonalizations (ED)\cite{usedpackage} for the $S=1/2$ (black line), $S=1$ (gray line) and $S=3/2$ (dashed gray line) on 12-site clusters.  Details are given in Appendices \ref{sec:LT} and \ref{sec:ED}  for (i) and (ii) respectively. 

\begin{figure}[ht!] \centering
  \includegraphics[clip,width=0.3\textwidth]{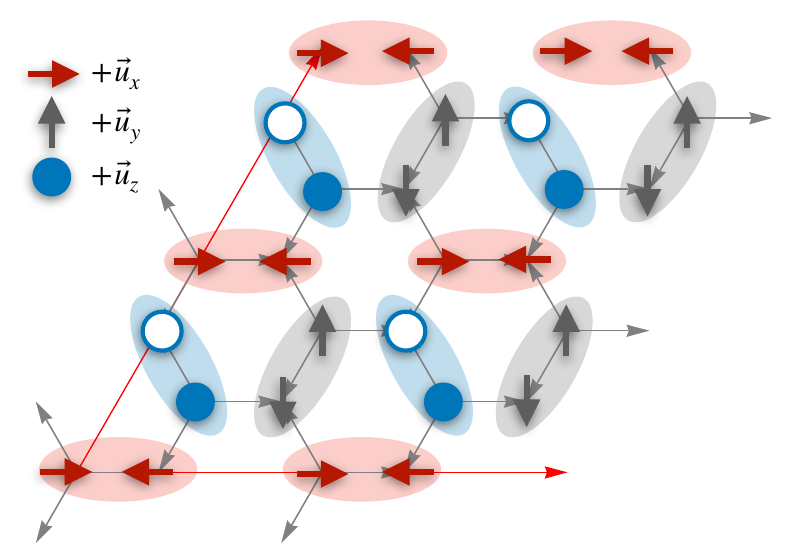}
  \caption{Sketch of the spin structure of a typical classical cartesian BSS state as obtained self-consistently at the Kitaev point $\theta=1/2$ and for a 24-site unit-cell. While obtained at $S=0.1$, the spin averages $\langle S_i^{x,y,z}\rangle $ follow the dimer star pattern. The classical energy per site $N$ and spin $S$ is obtained from ${\cal H}_\text{HK}$ as explained in the text. Classical dimers are represented by ovals, and the red, gray and blue colors correspond to spin orientations along the $x,y$ and $z$ axes. Dimers are ordered in a star pattern. 
  }
  \label{fig:bss}
\end{figure}

We see that our SBMFT construction at large $S$ matches perfectly the LT results, validating the proposed mixed singlet / triplet construction rather than the separated channels\cite{kargarian_unusual_2012,kos_quantum_2017,schneider_projective_2022} that would lead to a completely different ground state 
energy dependence on $\theta$. 
This systematic comparisons with the classical energies, as done here, provides crucial information for deriving the
appropriate Schwinger boson mean-field decoupling of the spin interactions. 
At the special Kitaev point $\theta = {\pi / 2}$, the classical state is known to be the Baskaran-Shen-Shankar (BSS) state made of fully-packed classical dimers distributed in a star pattern shape maximizing the number of empty hexagons\cite{baskaran_spin-_2008}. Every dimer possesses an internal degree of freedom -- its orientation --  hence resulting to a massive degeneracy.  Such a typical classical spin configurations can be obtained within SBMFT, properly recovering the expected classical energy per site $N$, ${E  \over N S^2} = -1$  is  by replacing $\langle {\bf S}_i  {\bf S}_j \rangle \to \langle {\bf S}_i \rangle \langle {\bf S}_j \rangle$ as expected in a classical situation with no quantum fluctuations, in ${\cal H}_\text{HK}$ and where $\langle {\bf S}_i \rangle$ is evaluated in the boson vacuum. An example of a typical classical dimer configuration is given in Fig.\ref{fig:bss}.

At large $S$, we have obtained the right sequence of phases with increasing $\theta$ as originally reported\cite{chaloupka_zigzag_2013} {\it i.e.} the Néel, 
the Zig-zag, the FM and the Stripy phases.
In Fig.~\ref{fig:energy}, we use filled symbols for the {\it Ans\"atze} that are accessible within the self-consistency, and empty ones for solutions that require to properly take into account the Bose condensation in the SBMFT. While it is not always necessary at small $S$ on finite systems since the spectrum is always gapped, it becomes unavoidable at larger $S$ for getting solutions satisfying the boson constraint. For example, the non-fluctuating ferromagnetic state at $\theta =\pi$ has all the bosons condensed at the $\Gamma$ point of the Brillouin zone, and a large portion of them remains condensed even at smaller $S$. In the rest of this paper, we focus on the regions of filled symbols where a good saddle point of the self-consistent equations can be stabilized.

A remarkable feature of our SBMFT is that under the lowering of $S$ down to $S=1/2$ and $S=1$, the GS energies remain very close to the 
ED energies (see Fig.~\ref{fig:energy}). However, the smaller the $S$, the larger the difference. This can be either attributed to the mean-field treatment, the finite size cluster used for the ED or even the fact that for the lowest spin, $S=1/2$, Majorana fermion excitations have significant effects close to the Kitaev point that cannot be described in the present theory. In this work, our main aim is to understand the nature of possible bosonic QSLs arising in the integer-spin HK model. Thus, in the rest of the paper we will focus on the properties, physical signatures, and robustness with $S$ of the QSLs found within our SBMFT. Our results are directly 
relevant to the characterization of the QSLs found numerically in other studies of the $S=1$ HK model \cite{dong_spin-1_2020,lee_tensor_2020}.

\section{Quantum spin liquids and chirality}
\label{sec:qsl}

In 2010, Wang  applied the projective symmetry group (PSG) method to classify all the possible $\mathbb{Z}_2$ spin liquid states of a SBMFT on the honeycomb antiferromagnet based on the ${\hat{h}^0}$ and ${\hat{p}^0}$ operators, up to third nearest neighbors.\cite{wang_schwinger_2010} It has been shown that 
only two QSLs are possible which can be distinguished by their gauge-invariant flux piercing the hexagonal plaquettes. The physical observable to quantitatively differentiate such non-trivial orders can be quantified through the Wilson loop (WL)\cite{tchernyshyov_flux_2006,wang_schwinger_2010}, defined as the phase $\phi_s= p^0_{ij} (-p^{0*}_{jk})p^0_{kl} (-p^{0*}_{lm})p^0_{mn} (-p^{0*}_{ni})$ around the six sites $i,j,k,l,m,n$ of an hexagonal plaquette. As Wang noticed\cite{wang_schwinger_2010}, in his construction of the PSG, the two QSL fluxes can be either $0$ or $\pi$, and both satisfy time-reversal symmetry (TRS). In general, it is however not forbidden to find chiral QSLs breaking TRS, as Messio {\it et al.} \cite{messio_time_2013}  have shown with a more general PSG treatment, on triangular and square lattices though. In our HK model, in addition to singlet flux $\phi_s$, we also define triplet flux as $\phi_t= p^1_{ij} (-p^{2*}_{jk})p^3_{kl} (-p^{1*}_{lm})p^2_{mn} (-p^{3*}_{ni})$. Other WL can be constructed, but $\phi_s$ and $\phi_t$ are enough to characterize all QSLs found in the present study.

\begin{figure}[ht!] \centering
  \includegraphics[clip,width=0.45\textwidth]{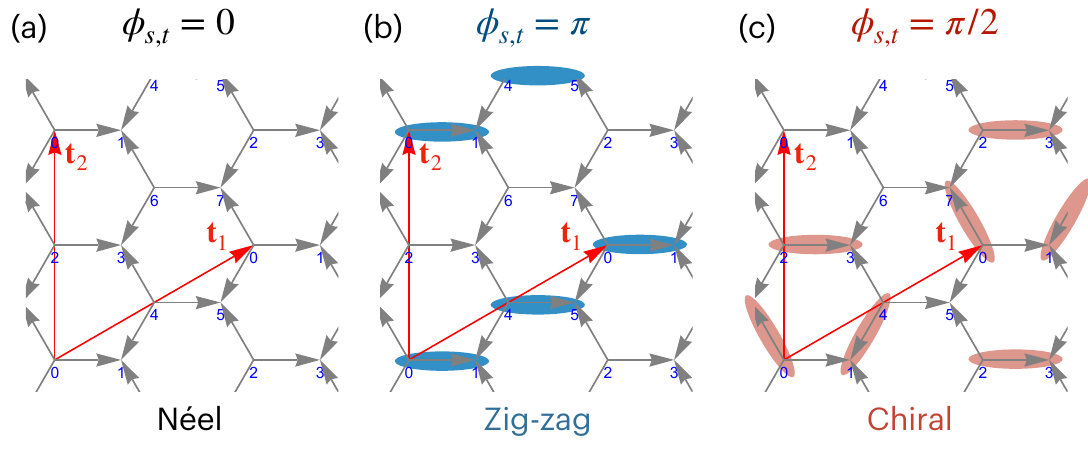}
  \caption{The eight-site magnetic unit cell considered in the text with its corresponding ${\bf t}_{1,2}$ translation vectors. Site indices are written in blue. In the SBMFT, each bond is oriented, here according to the convention indicated by the arrows. Thick bonds are extra phase that has to be multiplied to the mean field parameters, $-1$ (blue) or $i$ (red). Three QSL flux sectors are displayed, (a) the $\phi_{s,t} =0$ 
  with $n_u=2$, (b) the  $\phi_{s,t} =\pi$ with $n_u=4$ and (c) the chiral $\phi_{s,t} =\pi/2$ with $n_u=8$.
  }
  \label{fig:muc}
\end{figure}

\begin{figure}[ht!] \centering
  \includegraphics[clip,width=0.4\textwidth]{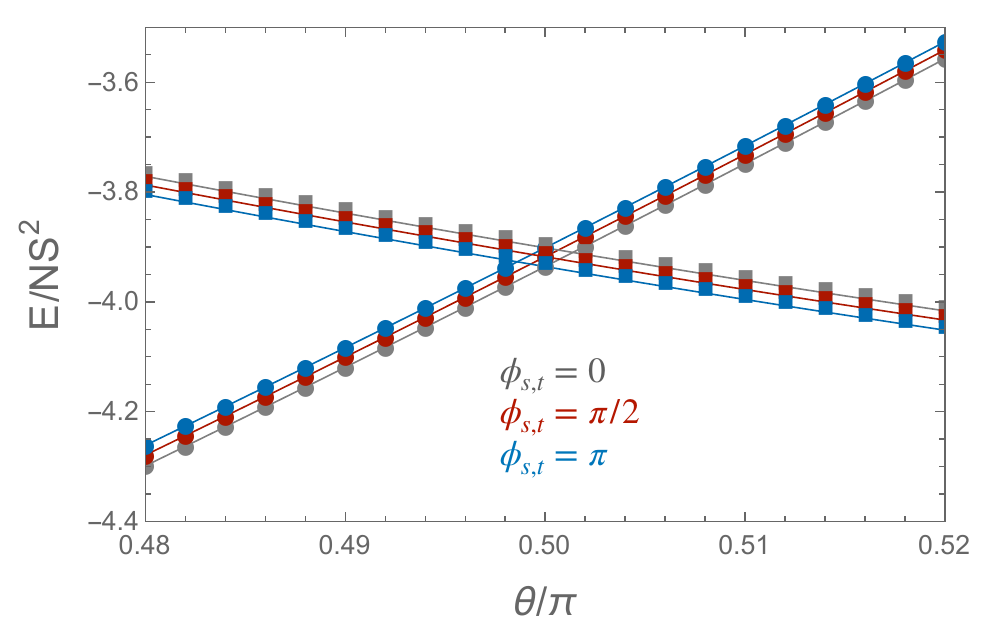}
  \caption{Zoom on the QSL energies around the pure Kitaev point at $\theta = \pi /2$ and at strong quantum fluctuations  ($S=0.1$) obtained by solving the SBMFT self-consistent equations. Each color corresponds to a flux sector: $0$ (black), $\pi/2$ (red) and $\pi$ (gray). Symbols correspond to the constituents of the QSLs, either singlets (circles) or triplets (squares). Exact degeneracies appear at $\theta = \pi /2$. States with flux of $\pi/2$ (red) are chiral.
  }
  \label{fig:qsls}
\end{figure}

\begin{table}
\begin{tabular}{|c|c|c|c|c|c|}
    \hline
     & $|p^0|$ & $|p^1|$ & $|p^2|$ &  $|p^3|$ & $\lambda$ \\
     \hline 
     $\phi_s=0$  &  0.1983(8) & 0 & 0 & 0 & 0.2293(8) \\
     $\phi_t=0$  &  $0$ & 0.1975(2) & 0.1975(2) & 0.1975(2) & 0.2240(3)\\
     $\phi_s=\pi/2$  &  0.1979(3) & 0 & 0 & 0 &0.2265(1)\\
     $\phi_t=\pi/2$  &  0 & 0.1979(3) & 0.1979(3) & 0.1979(3) & 0.2265(1)\\
     $\phi_s=\pi$  &  0.1975(2) & 0 & 0 & 0 & 0.2240(3)\\
     $\phi_t=\pi$  & 0 &  0.1983(8)  & 0.1983(8)  & 0.1983(8)  & 0.2293(8)  \\
     \hline
\end{tabular}
\caption{Non-zero mean-field parameters of the {\it Ans\"atze} at the Kitaev point $\theta = \pi/2$ and the corresponding Lagrange multipliers $\lambda$ for $S=0.1$. Note that at this specific point, exact degeneracy occurs between pairs of states, as can be seen in the energy zoom of Fig.~\ref{fig:qsls}. The results are obtained for a cluster of $n_u=8$ sites per unit cell and a linear size of $l=24$. }
\label{tab:param}
\end{table}
In a translationally and rotationally invariant triplet state, we  expect $p^\gamma_{ij} = p$.\cite{kos_quantum_2017} To focus on the properties of the QSLs, we first start our analysis for strong quantum fluctuations by reducing the value of the spin down to $S=0.1$. 
Solving the self-consistent equations, we identify 6 gapped phases, 3 made only of singlets ${\bf p}^0 \ne 0$ with $\phi_s \ne 0$ and $\phi_t$ not defined, and 3 made only of triplets ${\bf p}^\gamma \ne 0$ with $\phi_s \ne 0$ and $\phi_t$ not defined. 
In Table.~\ref{tab:param}, we give the explicit values of the non-zero mean field parameters at the Kitaev point $\theta = \pi /2$ and for a system with linear size  $l=24$ of $n_u=8$ unit-cells, and the corresponding phase structure is given in Fig.~\ref{fig:muc}.
Before going in more details in their physical properties, we  discuss in the rest of this section their energies, as shown in Fig.~\ref{fig:qsls} where circles correspond to the singlet, squares to the triplets and the colors gray, green, and blue are the corresponding fluxes  $\phi_{s,t} = 0$, $\pi/2$ and $\pi$ flux sectors. 

Note that the 2 singlet {\it Ans\"atze} at $\phi_s = 0$ and $\pi$ are the QSLs found by Wang.\cite{wang_schwinger_2010} They have  respectively a magnetic unit-cell of 2 and 4 sites, as shown in panels (a) and (b) of Fig.~\ref{fig:muc}. The lowest state is $\phi_s = 0$ that, by increasing $S$ (lowering the quantum fluctuations), eventually condenses to the Néel order as expected in the region $\theta < \pi /2$ (the complete analysis in function of $S$ is treated in the next section).  Interestingly, we also find a chiral 8-site magnetic unit-cell phase displayed on panel (c), with a flux of $\phi_s = \pi/2$. Its energy is in between the two other QSLs, without ever crossing their energies. This flux gives a pure imaginary phase of $i$ on the {\it Ansatz}, and thus breaks TRS 
implying a chiral QSL.   

Interestingly, applying the same phase structures to  triplet {\it Ans\"atze}, we find the 3 triplet counterpart of the 3 singlet liquids. The lowest triplet state is now in the $\phi_t = \pi$ sector, and it condenses to the Zig-zag state at large $S$ and in the region $\theta > 1/2$, as expected (see also Fig.~\ref{fig:energy}). As for the singlet states, we find a second chiral QSL with a flux $\phi_t = \pi / 2$. 

The energies of the singlet and triplet QSLs  cross each other at the pure Kitaev point $\theta = {\pi / 2}$ where every flux sector is  exactly doubly degenerate.  As previously mentioned, for such strongly frustrated model, the classical state is highly degenerate and an order by disorder mechanism
reduces the degeneracy to that of the BSS configurations\cite{baskaran_spin-_2008,samarakoon_comprehensive_2017,rousochatzakis_quantum_2018}. Quantum mechanically, spin liquids are found for $S=1/2$ and $S=1$ as recently reported in [\onlinecite{khait_characterizing_2021,hickey_field-driven_2020,zhu_magnetic_2020,dong_spin-1_2020,lee_tensor_2020}]. Our ED analysis shows how spin correlations, $\langle S^x_0 S^x_j \rangle$, (with $0$ a reference site) { drop  to} zero beyond the nearest-neighbor distance in the Kitaev model for $S=1/2, 1$ and 
$3/2$, the dependence of the spin correlations with intersite distance is qualitatively the same in the three cases (see Fig.~\ref{fig:gap} {of  Appendix B}). This indicates that indeed a QSL is present 
in the Kitaev model even for $S=3/2$ whose ground state energy is still quite below the LT energy (see Fig.~\ref{fig:energy}). We have explored further the nature of these QSLs by comparing the lowest excitation energies, $\Delta=E_1-E_0$, of the $S=1/2, 1$ Kitaev models 
using ED. A straightforward linear extrapolation of $\Delta$ calculated on the small clusters , $N \leq 24$, to the thermodynamic limit, suggests that the $S=1$ QSL is gapless as in the
the $S=1/2$ model which is known to be gapless from Kitaev's exact solution.  However, the limited cluster sizes available for the finite-size scaling analysis does not allow to conclude about 
the possible existence of a small gap in the $S=1$ QSL. {In fact while tensor network calculations\cite{lee_tensor_2020} indicate a gapped QSL for $S=1$ HK model, DMRG
concludes that a gapless QSL is present\cite{dong_spin-1_2020}.} Further calculations on larger systems are needed to clarify this important issue.

It is most likely that one of the present QSLs could be stabilized at larger $S$ when we go beyond the present mean field theory {as supported by the comparison with ED calculations, even though at the pure mean-field level, the gap is enlarged by increasing $S$}. Even beyond the scope of the present work, it would be 
 interesting to perform a complete PSG analysis for the present mean field construction in order to establish the list of all possible chiral {\it Ans\"atze}. We let this question open for a future work.
 
\begin{widetext}
\begin{center}
\begin{figure}[!ht]
  \includegraphics[clip,width=1.0\textwidth]{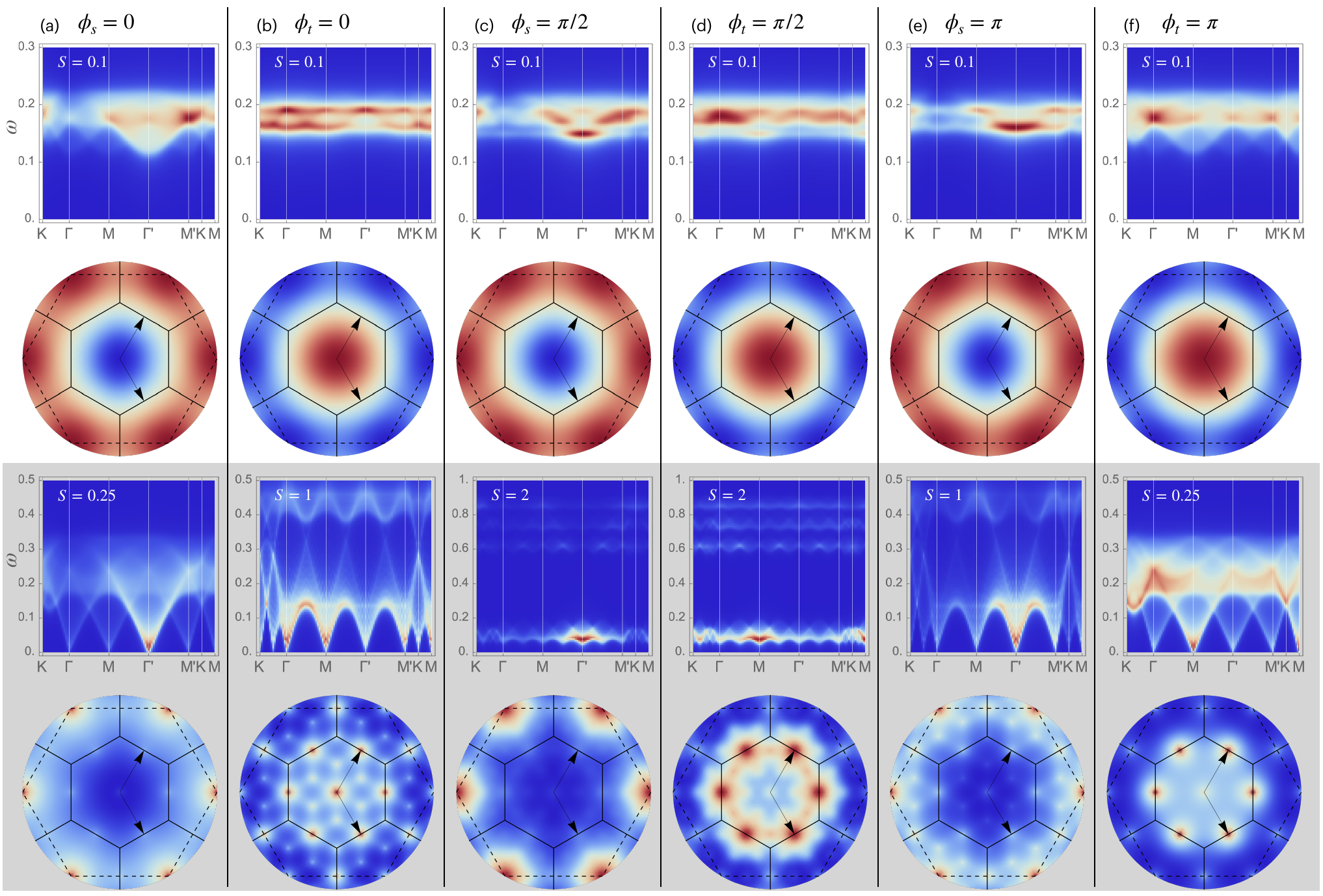}
  \caption{Dynamical and static spin structure factors for the QSLs discussed in the text., ordered according to their  flux$(\phi_{s,t}$ displayed at the top. All the results are obtained for a system of $n_u \times 24 \times 24$ sites with $n_u = 8$, at the Kitaev point $\theta = 1/2$. 
  {The six QSLs are ordered in columns, for which the}
  first two lines at the top are results obtained in the strong quantum fluctuation regime at $S=0.1$ where all phases are gapped. For the two last lines, {in the gray shadded box}, the same quantities are computed at a spin value  $S$ very close to the phase transition where the Bose condensation occurs for the corresponding {\it Ansatz}. The chiral states at $\phi_{s,t} = \pi /2$ are very stable upon increasing $S$ and remain a gapped spin liquid up to approximately a critical $S_c$ of $2.1-2.2$. Both dynamical and static structure factors provide clear distinct signatures for any of these phases. }  
  \label{fig:struct}
\end{figure}
\end{center}
\end{widetext}

\section{Physical signatures}
\label{sec:phys}

\begin{figure}[ht!]
  \includegraphics[clip,width=0.45\textwidth]{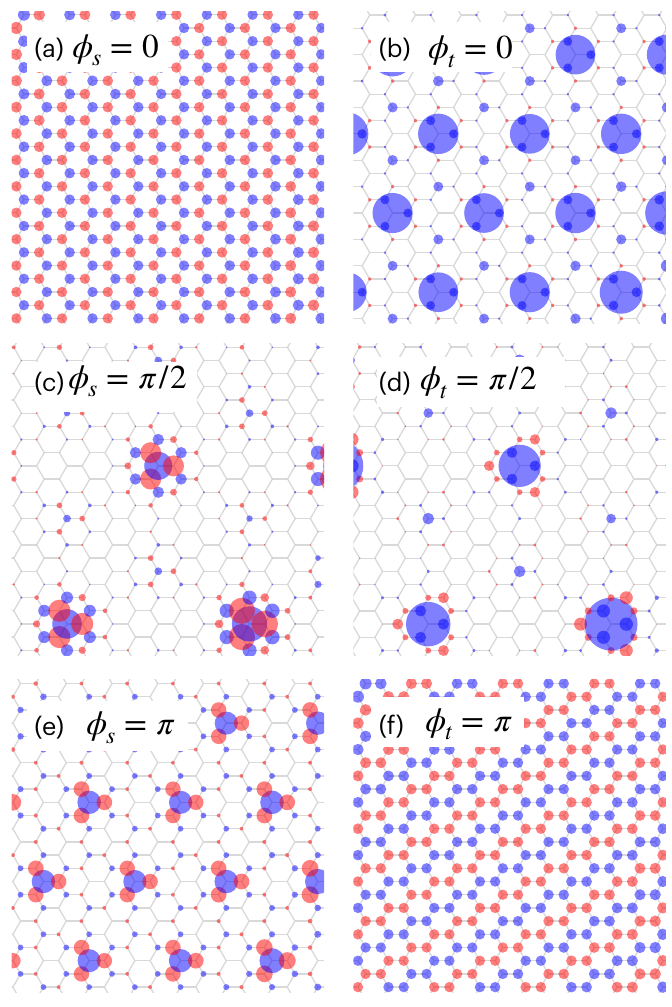}
  \caption{Real space spin-spin correlations $\langle {\bf S}_i \cdot {\bf S}_j \rangle$ for the 6 {\it Ans\"atze} discussed in the text. They are calculated deep in their crystalline regions  where the bosons are  already condensed, {\it i.e.} (a) and (f) at $S=1/2$, (b) and (e) at $S=2$ and (c) and (d) at $S=4$, for a cluster of $8\times 12 \times 12$ sites. Blue (red) disks correspond to positive (negative) correlations. The larger the radius, the stronger the correlations, in arbitrary units.}  
  \label{fig:real}
\end{figure}

It is very useful to characterize the QSLs found through their physical signatures such as dynamical and static structure factors. 
Also the critical $S_c$ at which they Bose condense provides interesting information of the QSL pattern. These questions are addressed 
in this section and in addition to previous gauge invariant Wilson loop observables, we consider the inelastic 
structure factor: 
\begin{eqnarray} 
	S({\bf q},\omega) = \frac{1}{N} \sum_{m,n} e^{i {\bf q} ({\bf r}_m - {\bf r}_n )} \int_{-\infty}^{+\infty} dt e^{-i \omega t} \langle { \bf S}_m \cdot { \bf S}_n \rangle,
\end{eqnarray} 
where $m,n$ run over all $N$ sites of the lattice. Details about the derivation can be obtained in the literature, {\it e.g.} in [\onlinecite{halimeh_spin_2016}] for an explicit expression in the case of the common $(\hat{\bf h}_0,\hat{\bf p}_0)$ SBMFT. Integrating over all frequencies $\omega$ leads to the equal-time structure factor $S({\bf q})$. Finally, together with the spin expectation values $\langle  {\bf S}_i\rangle$, we consider the real space spin-spin correlations to see magnetically ordered phases.
In Fig.~\ref{fig:struct} we show the dynamical and static spin structure factors for the QSLs, at $S=0.1$ (two top rows) and at $S \lesssim S_c$ (two bottom rows) . All results are obtained for the Kitaev model, $\theta = {\pi / 2}$ and for a system with $n_u = 8$ sites per unit cell, and linear size of $l=24$, namely with $N = 4608$ sites. In the following, we describe the dependence of these QSLs solutions on $S$ and their fate with the suppression of quantum fluctuations when increasing $S$ {when they eventually Bose condense to their corresponding  magnetically ordered states at higher but close values (gray shadded box), expected for the chiral QSLs that are still uncondensed up to $S=2$}. We have labeled all the phases (a) to (f) according to the order displayed in both Figs.~\ref{fig:struct} and \ref{fig:real}. This is also corroborated by their complex static structure factors, displaying distinctive experimental signatures  in actual compounds realizing the HK model.\\
{\it (a) {\it Ansatz} $\phi_s=0$:} As said, this state corresponds to the 0-flux $\mathbb{Z}_2$ QSL reported by Wang.\cite{wang_schwinger_2010} This fragile liquid is gapped for $S < 0.25$, and eventually quickly condenses to the Néel state above. At the Kitaev point, a level crossing with state $\phi_t = \pi$ (detailed below) occurs, and the Néel state is not the GS anymore. The structure factors show clear weights on the $\Gamma'$ points, in the second BZ.  This is illustrated in Fig.~\ref{fig:real} where we have calculated the real-space spin-spin correlations  $\langle {\bf S}_i  {\bf S}_j \rangle$. \\
{\it (b)  {\it Ansatz} $\phi_t=0$:} This state can be viewed as the triplet counterpart of the previous  {\it Ansatz}. However, one can see in Fig.~\ref{fig:struct} 
how at low $S$, the spectral weight in the dynamical structure factor is relatively flat and that Bose condensation occurs at a much higher spin value $S=1$  with different soft modes. In fact, the main Bragg peaks appear at the $\Gamma$ and $M$ points, showing the multi-$Q$ nature of the phase. Interestingly, the ordered magnetic state, arising upon Bose condensation, breaks translation symmetry leading to a corresponding large magnetic unit cell, 
as depicted in panel (b) of  Fig.~\ref{fig:real}. The structure is quite complex, with strong ferromagnetic correlations, isolated by regions of 
negative ones and surrounded by pins.\\ 
{\it (c)  {\it Ansatz} $\phi_s=\pi/2$:} With its pure imaginary Wilson phase, this state is chiral and breaks TRS. The gap is large at small $S$, and the liquid phase survives in a very large region of $S$, up to $S=2$. This is, together with phase (d) discussed just after, the best candidate for a bosonic QSL to be encountered in the HKM. This is unfortunately not the smallest solutions we find in the phase diagram, but its close proximity with the other solutions 
makes it a good candidate for the QSL found in the numerical studies of the $S=1$ Kitaev model \cite{dong_spin-1_2020,lee_tensor_2020}.  At very large spin $S=2$, the Bose condensation of the spinons leads to a very complex phase with a very large magnetic unit cell as depicted in panel (c) of Fig.~\ref{fig:real}. However, this state consisting only on singlets is certainly not the most exotic QSL found in the present work, despite its chirality.  On the other hand
and, as discussed below, the $S({\bf q}, \omega)$ of this QSL solution is the most consistent with numerical work on the $S=1$ HK model. \\
{\it (d)  {\it Ansatz} $\phi_t=\pi/2$:} This is certainly the most interesting QSL encountered here. This is a pure triplet QSL which, due to its purely 
imaginary Wilson flux it is also imaginary. The physical signatures of such state shown in panels (d) of Fig.~\ref{fig:struct} are quite exotic, with a very complex static structure factor at $S_c$,  and a strong almost flat low frequency signal even at $S=2$, in $S({\bf q}, \omega)$. The underlying magnetically ordered state consists on a large magnetic unit cell with very complex real space spin ordering (see Fig.~\ref{fig:real}(d)). It is interesting to note that the spins remain almost disordered even at large $S=2$, with local magnetic droplets well separated. This original feature is cleary different from the fermionic chiral QSLs found by the authors on the same model for the $S=1/2$ case with a Majorana mean-field theory\cite{ralko_novel_2020} and would be easily checked experimentally if such a state is present in actual materials. \\
{\it (e) {\it Ansatz} $\phi_s=\pi$:} This QSL is the second one reported by Wang in his 2010 article\cite{wang_schwinger_2010}. It has a $\pi$-flux per hexagon, and is more robust against the increase of $S$, since its Bose condensation occurs near $S=1$ (see Fig.~\ref{fig:struct}). It is worth noticing that the spin pattern obtained by spinon condensation was left as an open question by the author\cite{wang_schwinger_2010}. However, we clearly see in Fig.~\ref{fig:real} that, as the author speculated, the corresponding crystal phase is indeed very complex, with a large size magnetic unit cell, and various ferro- antiferro-magnetic correlations. \\
{\it (f)  {\it Ansatz} $\phi_t=\pi$:} In fact, the main Bragg peaks appear at the $M$ points, showing the multi-$Q$ nature of the phase. The ordered magnetic state upon Bose condensation is nothing else but the expected zig-zag phase, as shown by the real-space spin-spin correlations in panel (f) of Fig.~\ref{fig:real}, calculated for $S=1/2$. Our SBMFT provides qualitatively the correct sequence of magnetic phases.

An important question is which of the relevant SBMFT {\it Ans\"atze} discussed above is most likely to be realized in the
Kitaev model, $J_H=0$ with $S>{1 / 2}$ since there is no exact solution available for such models. 
Insight to this question can be gained by comparing the dynamical structure factor, $S({\bf q}, \omega)$, in 
Fig.~\ref{fig:struct} for the different SBMFT {\it Ans\"atze} with the
ED results of Fig.~\ref{fig:Sq}. An important qualitative feature of $S({\bf q}, \omega)$ is that the
AFM Kitaev model, $J_K>0$, has no spectral weight at the ${\bf q}=\Gamma$ point
not only for $S=1/2$ but also for $S=1, 3/2$ consistent also with the classical limit results\cite{samarakoon_comprehensive_2017,pohle_spin_2023}.
This is in contrast to the FM Kitaev model, $J_K<0$, where the $\Gamma$-point does have substantial 
spectral weight. A first inspection of $S({\bf q}, \omega)$ in Fig.~\ref{fig:struct} on the AFM Kitaev model 
shows how only the singlet {\it Ans\"atze} display spectral weight at the $\Gamma$-point. This is in consistent
agreement with the ED results for the $J_K>0$ of Fig.~\ref{fig:Sq}. On the 
other hand the dispersion spin excitations in the singlet QSL solution displayed in the upper row of 
Fig.~\ref{fig:struct}, for $S=0.1$, is rather flat around $\omega=0.18$. Hence both the presence of spectral weight around $\Gamma$ 
and the flat dispersion are in consistent agreement with the ED results for the $J_K>0$ of Fig.~\ref{fig:Sq}. 
Apart from this the singlet chiral phase with $\phi_s={\pi / 2}$ is characterized by a flat dispersion whose 
location is shifted to lower energies as $S$ is increased prevailing 
all the way up to $S \leq 2$. Such flat excitation dispersion at low energies is reminiscent of the dispersions found in 
the classical AFM Kitaev model\cite{samarakoon_comprehensive_2017}.

\section{Conclusions}
\label{sec:concl}

Motivated by the possibility of finding a bosonic Kitaev spin liquid in the integer spin-$S$ HK model, we have performed a detailed analysis 
applying a Schwinger boson approach on the model for arbitrary $S$ complemented with ED on the $S=1/2, 1
$ and $3/2$ cases, as well as the LTA valid in the large-$S$ classical regime. The latter has helped us to construct a proper 
mean-field theory that reproduces exactly the classical limit, as it should. This is achieved by rewriting the spin bilinears of the HK model 
in the bond eigenbasis made of singlet and triplet bonds, by introducing corresponding hopping and pairing operators. We have then focused on the possible $\mathbb{Z}_2$ QSLs in the strong quantum regime (very low $S$) and showed that the two already reported singlet QSLs\cite{wang_schwinger_2010} possess their triplet counterparts. Their physical signatures, the dynamic and static structure factors show clear distinct features, and are 
quite robust upon Bose condensation (increasing of $S$). Surprisingly, by enlarging the unit-cell to $n_u=8$ sites, we have been able to unravel, to the best of our knowledge, two novel chiral QSLs with a flux of $\pi/2$ piercing hexagons, one made of singlet and the second one its triplet counterpart. While bosonic chiral QSLS have been obtained within the projective symmetry group for the Kagomé and the triangular lattice\cite{bieri_projective_2016}, these phases have been overlooked on the honeycomb lattice so far because the projective symmetry group has only considered for non-chiral {\it Ans{\"a}tze}\cite{wang_schwinger_2010}. 
The fact that two chiral QSLs are found in the present work also opens the door for the search of other bosonic chiral QSLs.
Comparing our dynamics to the ones obtained with ED for $S=1/2$ and $S=1$, it seems that the ${\phi_s} ={\pi / 2}$ QSL found in the SBMFT 
is the most compatible QSL with the ED data. Interestingly, this QSL is very robust against increasing $S$ and remains gapped even up 
to $S=2$. This is a promising candidate for the intermediate QSL reported in previous works. We can expect that quantum fluctuations, 
beyond the SBMFT used here, will lower the energy of this $\phi_s={\pi / 2}$ below the rest of close-in-energy QSLs found making 
it the absolute ground state in a small but finite parameter regime of the HK model around the Kitaev point. The
consistency of the $S({\bf q}, \omega)$ of the $\phi_s={\pi / 2}$ QSL between SBMFT and ED suggests that
indeed this is the most probable candidate among the QSLs found within SBMFT for the ground state of the $S=1$ Kitaev model. 
Further numerical and/or analytical work beyond the SBMFT is needed to establish the definitive nature of QSLs in integer HK models.

\section*{Acknowledgment}
AR would like to thank  Laura Messio, Rico Pohle, Natalia Perkins and Harald O. Jeschke for stimulating discussions. {The authors would like to thank anonymous referees for usefull comments helping in improving this manuscript.} JM acknowledges financial support from (PID2022-139995NB-I00) 
MINECO/FEDER, Unión Europea and the Mar\'ia de Maeztu Program for Units of Excellence in R\&D
(Grant No. CEX2018- 000805-M).

\appendix

\section{Luttinger-Tisza approximation}
\label{sec:LT}

For completeness and illustrative purposes we briefly describe the Luttinger-Tisza approach\cite{luttinger_note_1951,litvin_luttinger-tisza_1974} 
to the Kitaev-Heisenberg model used previously \cite{kimchi_kitaev-heisenberg_2014,kishimoto_ground_2018,rousochatzakis_kitaev_2016}.
The LTA is a semiclassical approach 
to find the ground state energies of a given magnetic model. One introduces a Fourier transform of the 
spin operators: 
\begin{equation}
S^\alpha_{i,s} = {1 \over \sqrt{N_s}} \sum_{\bf q} e^{i {\bf k} \cdot {\bf R}_i } S^\alpha_{{\bf k}, s},
\end{equation}
where $R_i$ denotes the position of the unit cells in the honeycomb lattice (\ref{ham}), $s$ denotes the sublattice type and $N_s$ the 
number of sites on each sublattice. We here assume that there are only two sublattices: $s=A, B$ ($n_u=2$). 

The transformed Heisenberg-Kitaev model in ${\bf k}$ space 
then reads:
\begin{equation}
H= \sum_{{\bf k}, \gamma, m, n} S^\alpha_{{\bf k}, n} \Lambda_{nm}^\alpha({\bf k}) S^\alpha_{-{\bf k}, m} 
\end{equation}
where the three $2 \times 2$, $\Lambda^\alpha({\bf k})$, with the $\alpha=x,y, z$ matrices expressed as:
\begin{equation}
\Lambda^\alpha({\bf k})=
\begin{pmatrix}
0 & \Lambda^\alpha_{AB}({\bf k}) \\
 \Lambda^{\alpha*}_{AB}({\bf k}) & 0 \\
\end{pmatrix}
\end{equation}
 with:
 \begin{eqnarray*}
 \Lambda^x_{AB}({\bf k}) =\frac{1}{2} \left[ (2 J_K + J_H) e^{-i {\bf k} \cdot {\bf d}_1} + J_H e^{-i {\bf k} \cdot {\bf d}_2} + J_H e^{-i {\bf k} \cdot {\bf d}_3} \right],
 \nonumber \\ 
 \Lambda^y_{AB}({\bf k}) = \frac{1}{2} \left[ J_H e^{-i {\bf k} \cdot {\bf d}_1} +(2 J_K+ J_H) e^{-i {\bf k} \cdot {\bf d}_2} + J_H e^{-i {\bf k} \cdot {\bf d}_3} \right], 
 \nonumber \\
 \Lambda^z_{AB}({\bf k}) = \frac{1}{2} \left[ J_H e^{-i {\bf k} \cdot {\bf d}_1} + J_H e^{-i {\bf k} \cdot {\bf d}_2} + (2 J_K+J_H) e^{-i {\bf k} \cdot {\bf d}_3} \right],
 \end{eqnarray*}
 with ${\bf d_1}=(1,0)$, ${\bf d_2}=(-{1 \over 2},- {\sqrt{3} \over 2} )$,  ${\bf a_2}=(-{1 \over 2}, {\sqrt{3} \over 2} )$.
 The Luttinger-Tisza condition on the absolute spin magnitude of the whole lattice reads:
 \begin{equation}
 \sum_{{\bf k},n} {\bf S}_{{\bf k},n} 
 \cdot {\bf S}_{-{\bf k},n} = 2 N_s S^2=N S^2,   
 \end{equation}
with $n_u$ the number of unit cells in the lattice. The constraint is introduced through a single Lagrange 
multiplier, $\lambda$, in the free energy: $F=H-\lambda (\sum_{{\bf k},n} {\bf S}_{{\bf k},n} 
 \cdot {\bf S}_{-{\bf k},n} - N S^2 )$. The minimization of $F$ leads to a set of self-consistent 
 equations:
 \begin{equation}
 \sum_{m} \Lambda^\alpha_{nm}({\bf k}) S^\alpha_{{\bf k}, m} = \lambda S_{{\bf k}, n}^\alpha. 
 \end{equation}
 Hence, from the diagonalization of each $\Lambda^\alpha({\bf k})$ matrix,  we obtain a set of eigenvalues $\lambda$. 
 For a given eigenvalue, the energy of the system can be expressed as:
 \begin{eqnarray}
 H&=&\sum_{{\bf k}, \alpha, n}  \left( \sum_m  \Lambda^\alpha_{nm}({\bf k}) S^\alpha_{{\bf k}, m}  \right)  S^\alpha_{-{\bf k},n} 
 \nonumber \\
 &=& \lambda \sum_{{\bf k}, n} S^\alpha_{{\bf k}, n} S^\alpha_{-{\bf k}, n}=  \lambda S^2.
 \end{eqnarray}
 
So the energy per unit cell of the system is given by the lowest
 $\lambda$ common to all three $\Gamma^\alpha$ matrices. 
 The ground state energy is given by the lowest  $\lambda$ on 
the 1st Brillouin zone. 

An analytic expression of the eigenvalues of, say $\Lambda^x({\bf k})$, can be obtained and reads:
\begin{widetext}
\begin{eqnarray*}
\lambda_{\pm} ({\bf k} ) =\pm {1 \over 2} \sqrt{ (2 J_K +J_H)^2 + 2 J_H^2 + 2 (2 J_K+ J_H) J_H \cos ({\bf k} \cdot {\bf \xi}_1) + 
2 (2 J_K+ J_H) J_H \cos ({\bf k} \cdot {\bf \xi}_2 )+ 2 J_H^2 \cos ({\bf k} \cdot {\bf \xi}_3) },
\end{eqnarray*}
\end{widetext}
with ${\bf \xi}_1 = {\bf d}_1 - {\bf d}_2$, $\xi_2 = {\bf d}_1 - {\bf d}_3$ and $\xi_3 = {\bf d}_2 - {\bf d}_3$ expressed in terms of the nearest-neighbor 
vectors: ${\bf d}_1=(1,0)$, ${\bf d}_2=(-1/2,-\sqrt{3}/2))$  ${\bf d}_3=(-1/2,\sqrt{3}/2)$. 

\begin{figure}[ht!] \centering
  \includegraphics[clip,width=0.35\textwidth]{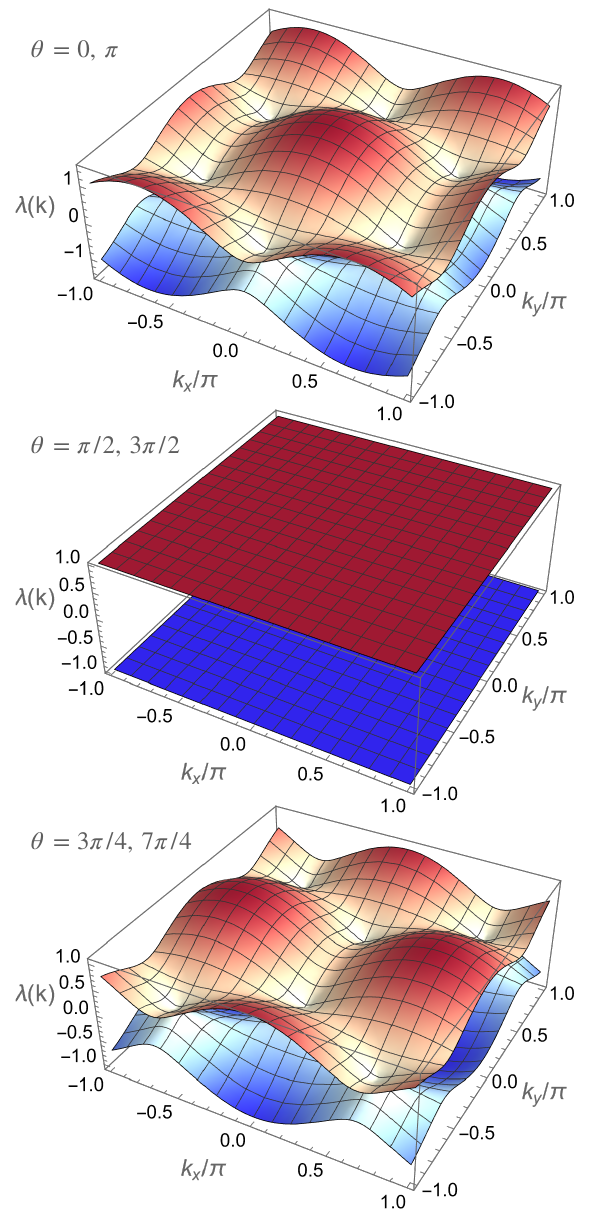}
  \caption{Eigenvalues obtained from the Luttinger-Tisza approach on the Kitaev-Heisenberg model.
  The ${\bf k}$ dependence of $\lambda_{\pm}({\bf k})$ eigenvalues are shown for the pure Heisenberg model, $\theta=0$,
  the pure Kitaev, $\theta= \pi/2$, the intermediate Heisenberg-Kitaev models, $\theta=3 \pi/4$ ($J_H<0, J_K>0$)
  and $\theta=7 \pi/4$ ($J_H>0, J_K<0$). }
  \label{fig:lambdaq}
\end{figure}

For illustrative purposes we show in Fig.~\ref{fig:lambdaq} the LT eigenvalues, $\lambda_{\pm} ({\bf k} )$ for
 three representative model parameters including the Heisenberg model, $J_K=0$, with $J_H>0$ ($\theta=0$) and $J_H<0$ 
 ($\theta=\pi$), the Kitaev model, $J_H=0$, with $J_K>0$ and $J_K<0$ and the intermediate $J_K, J_H \neq 0$ 
 Kitaev-Heisenberg models  $J_H<0, J_K=-2 J_H>0$ ($\theta=3 \pi/4$)  and  $J_H>0, J_K=-2 J_H<0$ ($\theta=7 \pi/8$).
Based on these results we consider the different cases. 

{\it Heisenberg models: $\theta=0, \pi$} 

As can be observed in Fig.~\ref{fig:lambdaq}, for both $\theta=0, \pi$, the lowest eigenvalue $\lambda_-({\bf k})$ 
has a minimum at ${\bf k}=\Gamma$ with value: $\lambda_-({\bf k})=\lambda(\Gamma) = -{3 / 2} J_H$. 
Hence, the ground state energy per unit cell is:
\begin{equation}
{E_0 \over N_s}=2 S^2 \lambda_-( \Gamma) = -3 S^2 J_H,
\end{equation}
for both $\theta=0, \pi$. 
For $\theta=0$, the ground state eigenvector is:
\begin{equation}
S^\alpha_{{\bf Q}=\Gamma, s}= (-1)^s,
\end{equation}
wit, say, $s=0$ for $A$ sites and $s=1$ for $B$ sites and $\alpha=x,y$ or $z$. Hence, the 
ground state is N\'eel ordered with a fully saturated staggered magnetic moment, $|S_{i, s}|^2 =S^2$. 

For $\theta=\pi$, the ground state spin order is:
\begin{equation}
S^\alpha_{{\bf Q}=\Gamma, s}= \pm 1,
\end{equation}
which is just a saturated FM  with $|S_{i, s}|^2 =S^2$. 

These results recover the expected ground states of the AF and FM Heisenberg models on 
the honeycomb lattice. The order can be along the $x$, $y$ or $z$ directions.

{\it Kitaev models: $\theta=\pi/2,3\pi/2$} 

In these cases the eigenvalues, $\lambda({\bf k})$, are independent of ${\bf k}$ leading
to the two flat bands showed in Fig.~\ref{fig:lambdaq}.
The lowest eigenvalue has a value:
\begin{equation}
\lambda_{\pm}({\bf k})=  \pm J_K.
\end{equation}
with a corresponding ground state energy per unit cell:
\begin{equation}
{E_0 \over N_s}=2 S^2 \lambda_-( \Gamma) = -2 S^2 J_K.
\end{equation}
The  ground state eigenvector is independent of ${\bf k}$:
\begin{equation}
S^\alpha_{\bf k}=(-1)^s,
\end{equation}
implying no preferred ordering of the spins: a classical spin liquid.
This is expected since the Kitaev model is known to lead to a 
spin liquid even in the classical model. The ground state energy per unit cell 
coincides with the previously found\cite{baskaran_spin-_2008}, ${E_0 \over N_s}=-J_K$ 
(note that in our model the Kitaev term is multiplied by a factor of 2), as it should.

{\it Heisenberg-Kitaev model: $\theta=3\pi/4, 7\pi/4$}

In the intermediate parameter regime of the Kitaev-Heisenberg model for both
$\theta=3\pi/4, 7 \pi/4$ the lowest eigenvalue has minima at the 
$M$-points. The ground state energy per unit cell is:
\begin{equation}
{E_0 \over N_s}=2 S^2 \lambda_-(M) = -2.13182 S^2 J_K.
\end{equation}

The ground state magnetic order for the $z$ spin component reads:
\begin{equation}
S^x_{i,s}= \cos({\bf Q} \cdot {\bf R}_i) (-1)^s,
\end{equation}
where ${\bf R}_i$ are Bravais lattice vectors. The corresponding ordering vectors, ${\bf Q}=M=(\pm {2 \pi \over 3},0),$
describe zig-zag magnetic 
order along the $z$-bonds, as reported previously. Zig-zag order along the $y$ ($z$) bonds is found from the
solution of $\Lambda^y({\bf k})$ ($\Lambda^z({\bf k}$) ) matrices, as it should.

In contrast for $\theta=7 \pi/4$, the ground state magnetic order reads:
\begin{equation}
S^x_{i,s}= \cos({\bf Q} \cdot {\bf R}_i),
\end{equation}
with, again, ${\bf Q}=(\pm {2 \pi \over 3},0)$ which describes stripy magnetic order along the
$x$-direction. Stripy order along the $y$ ($z$) bonds is found from diagonalizing the
$\Lambda^y({\bf k}$ ($\Lambda^z({\bf k}$) ) matrices, as it should.

\section{Exact diagonalization analysis}
\label{sec:ED}
Several works have treated the ground state properties of the Kitaev model for $S \ge 1/2$ with exact diagonalization
techniques \cite{usedpackage,koga_ground-state_2018,hickey_field-driven_2020,khait_characterizing_2021}. Here we provide 
our relevant ED calculations \cite{evenbly_improving_2014} performed to compare with the SBMFT of the $S \ge 1/2$ Kitaev model presented in the main text for completeness. 

\begin{figure}[ht!] \centering
  \includegraphics[clip,width=0.35\textwidth]{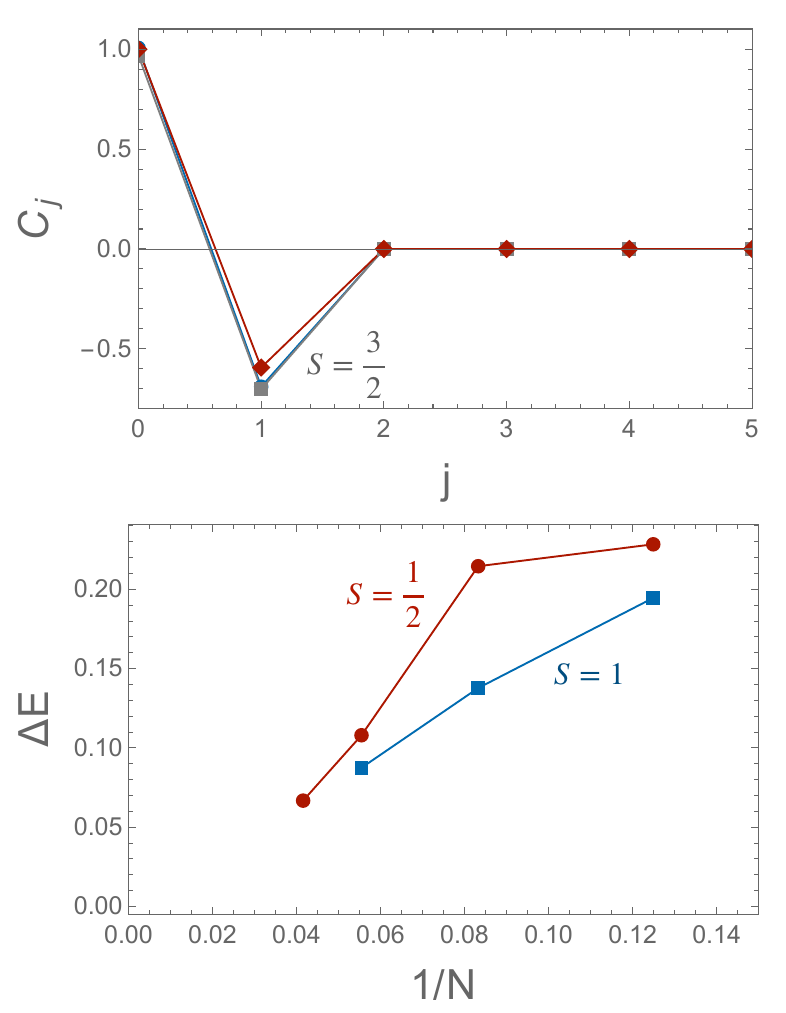}
  \caption{Normalized spin correlations $C_j = 3 \langle S_0^x S_j^x\rangle / S(S+1)$ and the gap in the $S \ge 1/2$ Kitaev model. The spin correlations, $\langle S^x_0 S^x_j \rangle$,
  in the ground state of the Kitaev model are compared for $S=1/2, 1, 3/2$. The dependence of the gap, $\Delta E=E_1-E_0$, 
  on $1/N$ of the $S=1/2$ Kitaev model, $J_H=0, J_K=2$, obtained from exact diagonalization up to $24$ sites 
  compared with the dependence of the $S=1$ Kitaev model.}
  \label{fig:gap}
\end{figure}

In order to explore the magnetic properties of the ground state of the pure Kitaev model, $J_H=0, J_K=2$, 
on $S$, we have calculated
the normalized real space spin correlations, $C_j = 3 \langle S_i^x S_j^x\rangle / S(S+1)$ with $i=0$ taken as an arbitrary site of the lattice as 
shown in Fig.~\ref{fig:gap}. The spin correlations are found to be non-zero only up to the nearest-neighbors 
for the $S=1/2, 1, 3/2$ cases explored. This indicates that the ground state is spin disordered and is a 
QSL up to $S=3/2$. This is a consequence of the conservation of the $\mathbb{Z}_2$ gauge fluxes 
around the hexagonal plaquettes\cite{baskaran_spin-_2008}:
 \begin{equation}
 W_p= e^{i \pi ( S_i^y + S_j^z + S_k^x+S_l^y+S_m^z+S_n^x)} 
 \label{eq:wp}
 \end{equation} 
where $i, j, k, l, m, n$ denote six sites around the plaquette, $p$, where the spin operators entering 
(\ref{eq:wp}) correspond to the bonds sticking out from each hexagonal vertex. Another important question is 
whether the Kitaev model for $S>1/2$ is gapped or not. In Fig.~\ref{fig:gap} we perform a finite size scaling of the 
excitation gap $\Delta E=E_1-E_0$ comparing 
the $S=1/2$ and $S=1$ Kitaev model. The gap is calculated in clusters with $N=8,12,18$ and 
24 sites (only for the $S=1/2$ case). A straightforward linear extrapolation of the gap obtained on the 
two largest clusters to the thermodynamic limit ($1/N \rightarrow 0$) would lead to a zero gap indicating
that the $S=1$ case is gapless as the $S=1/2$. However, due to the limitations on the 
size of the accessible clusters a definitive conclusion on whether the $S=1$
model is gapped or not cannot be reached. Our analysis seems to indicate a more rapid decrease of the 
 $S=1/2$ gap compared with the $S=1$ case. Further work on larger clusters is indeed 
 needed to reach a definitive conclusion on the size of the gap of the $S=1$ Kitaev model.

The physical spin excitations probed in inelastic neutron scattering experiments can be obtained from the exact
dynamical spin spectral function:
\begin{equation}
S({\bf q}, \omega)= \sum_\alpha \sum_n | \langle n | S^\alpha_{\bf q} |0 \rangle |^2 \delta( \omega-(E_n-E_0))  
\end{equation} 
where:
\begin{equation}
S^\alpha({\bf q}, \omega)={ 1 \over \sqrt{N} } \sum_m e^{i {\bf q} \cdot {\bf r}_m } S_m^\alpha,
\end{equation}
with $\alpha=x,y,z$ and ${ \bf r}_m$ denotes lattice site which can be alternatively
denoted by the position of the unit cell, ${\bf R}_i$, and the sublattice index as $({\bf R}_i, s)$.

\begin{figure}[ht!] \centering
  \includegraphics[clip,width=0.5\textwidth]{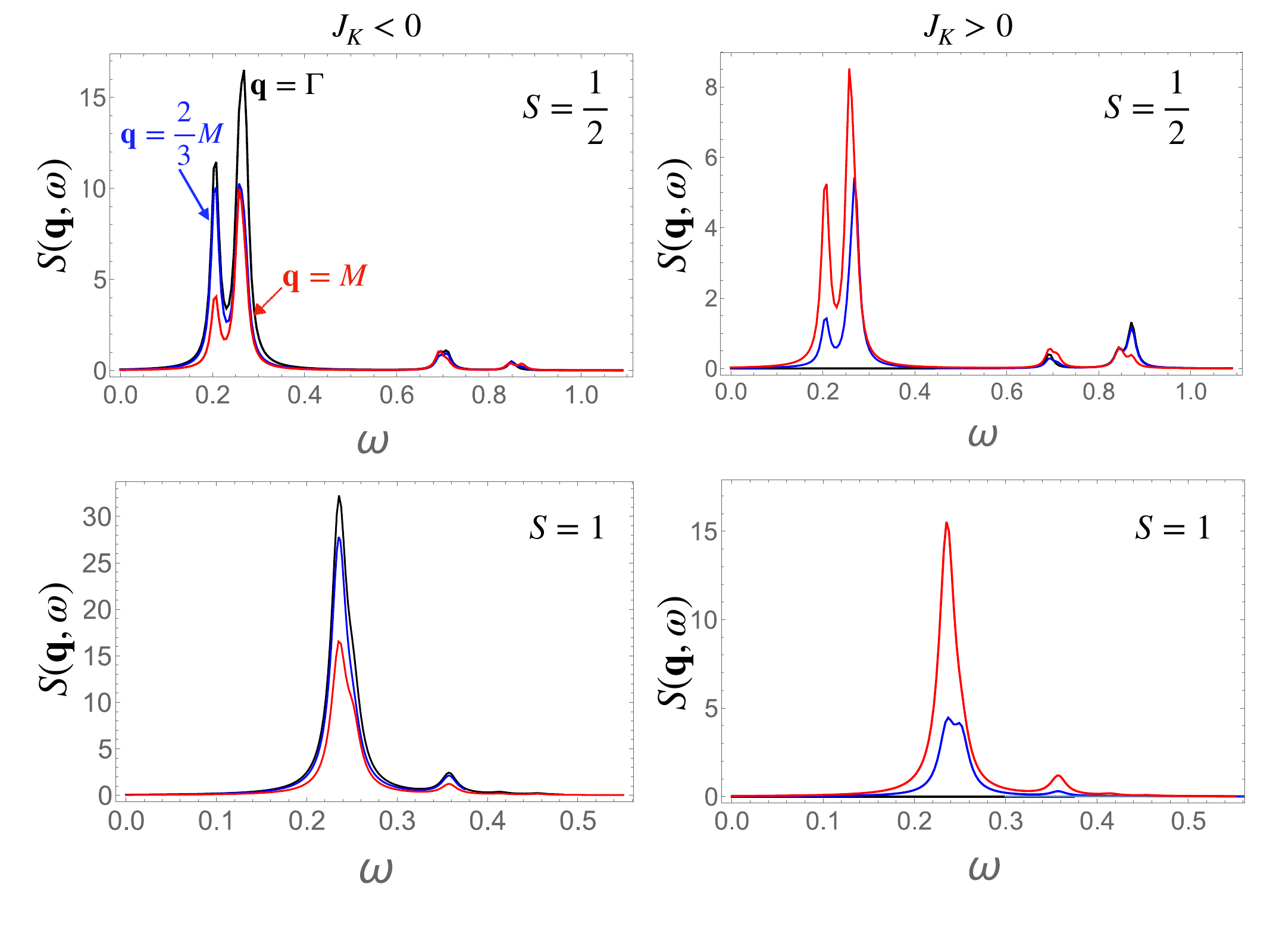}
  \caption{Dynamical spin spectral function of the Kitaev model. The $S({\bf q},\omega))$ of the $S=1/2$ Kitaev model 
  ($J_H=0$) is compared with the $S=1$ case for antiferromagnetic and ferromagnetic $J_K$. The characteristic flat 
  dispersion spectra of the $S=1/2$ Kitaev model is also present for $S=1$.  The figure shows the dispersion of the 
  spin excitations along the $\Gamma$-$M$ direction.}

  \label{fig:Sq}
\end{figure}

The $S({\bf q}, \omega)$ on the pure Kitaev model obtained 
with ED on $N=12$ site clusters is shown in Fig.~\ref{fig:Sq} along the $\Gamma-M$ direction. 
The $S({\bf q}, \omega)$ for $S=1/2$, displays a gap to a dispersionless band associated with
the localized vison excitations as previously found\cite{baskaran_spin-_2008,gohlke_dynamics_2017}. 
In agreement
with other numerical works, the antiferromagnetic, 
$J_K>0$, $S=1/2$ Kitaev model displays no weight at the ${\bf q}=0$ mode in 
contrast to the ferromagnetic model with $J_K<0$ as found previously \cite{samarakoon_comprehensive_2017,pohle_spin_2023}. 
Although the $S({\bf q}, \omega)$ for $S=1$ and $S=1/2$ are very similar there are also some
differences. For instance, the characteristic peaks around $\omega=0.25$ 
describing the dispersionless spin excitations of the Kitaev model 
are broader for $S=1/2$ than for $S=1$.  

\begin{figure}[ht!] \centering
  \includegraphics[clip,width=0.35\textwidth]{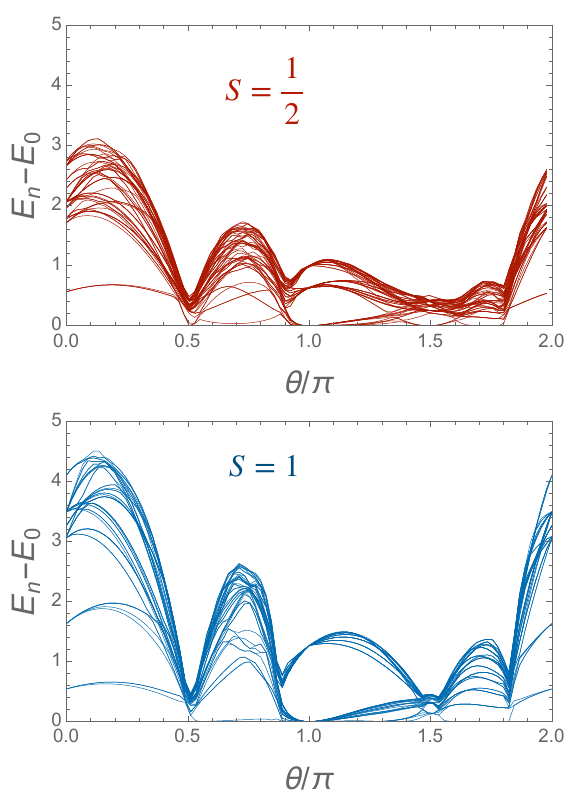}
  \caption{Excitation energy spectra  $E_n-E_0$ of the Heisenberg-Kitaev model. The dependence of the 
  excitation energies, $E_n-E_0$ on $\theta$  are shown for the $S=1/2$ and $S=1$  models.
  A dense level spectrum suggesting a closing of the gap of the $J_K>0$ Kitaev model $S=1/2$, 
  $\theta=\pi/2$, is found.
  }
  \label{fig:spectra}
\end{figure}

Further insight into the nature of the various magnetic states arising in 
the Heisenberg-Kitaev model can be gained from the direct inspection of its 
excitation spectrum. As discussed above, we are specially interested in knowing whether QSLs
arising in the $S>1/2$ Kitaev model are gapped or not. This can be explored 
by comparing the dependence of the energy excitation spectra on $\theta$
of the $S=1$ HK model with the $S=1/2$ case as shown in Fig.~\ref{fig:spectra}.
The figure displays a pronounced 
suppression of the excitation energies around $\theta=\pi/2$ {\it i.e.} for the 
$S=1$ AF Kitaev model. This behavior is also found in the $S=1/2$ case around 
$\theta=\pi/2$ (see Fig.~\ref{fig:spectra}) where Kitaev exact gapless QSL is expected. 
The suppression of excitation energies also arises around $\theta=3\pi/2$ {\it i.e} for the FM Kitaev model 
somewhat suppressed for $S=1$ compared to $S=1/2$. Due to the small cluster available ($N=12$ sites),
it is difficult to reach a definitive conclusion on whether the $S=1$ Kitaev model 
is gapped or not based solely on our ED results.
{Tensor network calculations \cite{lee_tensor_2020} do suggest that the $S=1$ Kitaev model is gapped. 
However, DMRG calculations \cite{dong_spin-1_2020} on cylinders are more consistent with a 
gapless QSL.} Further
numerical work is needed to establish this definitely. In any case, the possibility of a bosonic 
$\mathbb{Z}_2$ QSL as predicted here by SBMFT is not excluded by 
either our ED or previous numerical studies on larger systems. 

\bibliography{library}

\end{document}